\documentstyle[preprint,revtex]{aps}

\begin{document}
\thispagestyle{empty}


\begin{center}
\hfill{IP-ASTP-24-94}\\
\hfill{February 1995}\\
\hfill{(Revised version)}\\
\vspace{1 cm}

\begin{title}
Large-angle Polarization of the\\ Cosmic Microwave Background Radiation\\
and Reionization
\end{title}
\vspace{0.5 cm}

\author{Ka Lok Ng and Kin-Wang Ng}
\vspace{0.5 cm}

\begin{instit}
Institute of Physics, Academia Sinica\\ Taipei, Taiwan 115, R.O.C.
\end{instit}
\end{center}

\begin{abstract}
We discuss the effect of matter reionization on the large-angular-scale
anisotropy and polarization of the cosmic microwave background radiation
(CMBR) in the standard CDM model. We separate three cases in which the
anisotropy is induced
by pure scalar, pure tensor, and mixed metric perturbations respectively.
It is found that, if reionization occurs early enough, the polarization can
reach a detectable level of sequentially $6\%$, $9\%$, and $6.5\%$ of the
anisotropy. In general,
a higher degree of polarization implies a dominant contribution from the
tensor mode or reionization at high redshift.  Since early reionization will
suppress small-scale CMBR
anisotropies and polarizations significantly,
measuring the polarization on few degree scales
can be a direct probe of the reionization history of the early universe.

\vspace{0.5 cm}
\noindent
{\it Subject headings:} cosmology: cosmic microwave background --- cosmology:
observations
\end{abstract}
\newpage

\noindent
{\centerline{\bf 1. INTRODUCTION}}

\noindent

The large-scale anisotropy of the cosmic microwave background radiation
(CMBR)
measured by the DMR onboard the Cosmic Background Explorer satellite (Smoot
et al. 1992) may be induced by
density perturbations (scalar mode) and primordial gravitational waves
(tensor mode) via the Sachs-Wolfe (SW) effect (Sachs \& Wolfe 1967).
If the anisotropy is mainly due to the scalar mode, the COBE result combined
with observations of the large-scale structure would provide us with an
important clue for discriminating between different cosmological models.
However, it was recently argued that this anisotropy might
be dominated by the tensor mode (Krauss \& White 1992).  Furthermore, it
was shown that the
tensor-mode dominance actually occurs in certain inflation models
(Davis et al. 1992). Therefore, distangling the scalar from the tensor
contributions is not only very important for testing the inflation model but
is also needed to understand the formation of large-scale structure.
It was then suggested that by comparing large- and
small-scale anisotropy measurements, one can separate the scalar- and
tensor-mode contributions (Davis et al. 1992; Crittenden et al. 1993).

However, early reionization of the universe
may influence the formation of structures
(Couchman \& Rees 1986) as well as damp out small-scale CMBR anisotropies
(Vittorio \& Silk 1984; Bond \& Efstathiou 1984). The Gunn-Peterson test
indicates that the universe must have been highly reionized by a redshift of
five or greater (Gunn \& Peterson 1965). Recently, the likelihood of early
reionization by radiation emitted from young galaxies has been investigated in
detail. In a standard cold dark matter
(CDM) model, typical parameter values predict that reionization occurs at a
redshift around 50 (Tegmark, Silk, \& Blanchard 1994); similar results were
also obtained by numerical simulation (Fukugita \& Kawasaki 1994).
When normalized to COBE observations,
the CDM model very likely has reionization at redshifts around
28-69 (Liddle \& Lyth 1994). With reionization at redshift $\sim 50$,
the Doppler peak on degree scales may be damped almost completely away while
large-scale anisotropies remain reasonably unaffected
(Sugiyama, Silk, \& Vittorio 1993). Indeed, it was attempted to have early
reionization to smooth out excessive temperature fluctuations on degree
scales predicted in the CDM model, in
order to reconcile the model with the lowest limits from South-Pole 91 data
(Gaier et al. 1992). However, recent small-scale anisotropy measurements
from different groups have already hinted that the Doppler peak seems to be
present at around the correct height for models based on adiabatic density
perturbations without reionization.
For nonstandard CDM models, some of them could have reionization
occuring at redshifts early enough (Liddle \& Lyth 1994)
to suppress degree-scale anisotropies (Kamionkowski, Spergel, \& Sugiyama 1993;
Tegmark \& Silk 1994a). Therefore, future anisotropy measurements on degree and
subdegree angular scales would be crucial for determining the reionization
history of the universe.

Polarization of the CMBR is another clue that could have a great
potential of probing the reionization history of the universe.
Anisotropic radiation acquires linear polarization when it is scattered with
free electrons (Rees 1968). There have been several works on calculating
the small-angular-scale ($\le 1^o$) r.m.s. polarization of the CMBR induced by
adiabatic density perturbations in an universe with standard recombination
(Kaiser 1983; Bond \& Efstathiou 1984), and in a reionized universe
(Bond \& Efstathiou 1984; Nasel'skii \& Polnarev 1987).  It was shown that
roughly $10-20\%$ of the CMBR anisotropy is polarized on arc-minute scales.
A thorough numerical calculation for both large- and small-
angular-scale polarization of the CMBR induced by adiabatic and isocurvature
density perturbations with standard recombination has been performed (Bond \&
Efstathiou 1987). Their calculations have confirmed earlier small-scale results
and shown that large-scale polarization is insignificant.

An analytic estimation of the quadrupole polarization induced by scalar and
tensor metric perturbations was made in various cosmological models including
matter reionization (Ng \& Ng 1993). The r.m.s. temperature
anisotropy and polarization of the CMBR induced by the tensor mode
perturbation of arbitrary wavelength was also computed (Polnarev 1985; Frewin,
Polnarev, \& Coles 1993; Ng \& Ng 1994).
It was shown that large-scale polarization is greatly enhanced by
early reionization.

Similarly, the large-scale polarization of the CMBR induced by scalar and
tensor modes within inflationary models was investigated
under the assumption of no reionization (Harari \& Zaldarriaga 1993).
A detailed numerical calculation of scalar and tensor contributions to the
CMBR polarization power spectrum in inflationary models
has also been carried out (Crittenden, Davis, \& Steinhardt 1993). Their
calculations show that the polarization can reach a $10\%$ level of the
anisotropy in an universe with no hydrogen recombination, and may be
a useful discriminant for determining the ionization history of the universe.

In this paper, we will investigate in detail the CMBR polarization induced by
scalar and tensor modes in the presence of early reionization at redshift
around $30\sim 90$. In this
scenario, CMBR anisotropies on degree and subdegree scales are suppressed
significantly. Hence, although the degree of polarization is still about
$10\%$, the absolute polarization would be suppressed.
In contrary, large-scale anisotropies measured by COBE remain reasonably
unaffected while the degree of large-scale polarization is greatly enhanced.
Thus, measuring large-scale CMBR polarization would become more important.

In our previous paper (Ng \& Ng 1994), we have given a detailed numerical
calculation of the polarization of CMBR induced by
pure tensor modes in an universe with and without matter reionization.
It was shown that future polarization measurements with
windows for the power spectrum from $l=2$ to $l=50$, at sensitivity of
$10\%$ level of the anisotropy, might have a chance to detect the CMBR
polarization if early reionization took place at redshifts $\ge 90$.
Here we will combine these tensor-mode results with the
scalar-mode contribution.
The logic of this paper is more or less similar to the work by Crittenden
et al. (Crittenden, Davis, \& Steinhardt 1993). The difference is that we will
not pursue models with no recombination, since they are disfavored (Tegmark \&
Silk 1994b) by the COBE FIRAS data (Mather et al. 1994).

\noindent
{\centerline{\bf 2. METHODOLOGY}}

\noindent

We shall use the units $c=\hbar=1$ throughout.
Our calculations are based on the standard CDM model with a flat metric:
$ds^2=a^2(\eta) \left(d\eta^2-d{\bf x}^2 \right)$,
where $a(\eta)$ and $d\eta=dt/a(t)$ are the scale factor and conformal
time respectively. Here we normalize the conformal time to unity today. In this
metric, $\Omega_{\rm total}=\Omega_{CDM}+\Omega_B=1$,
where $\Omega_{CDM}$ and $\Omega_B$ denote respectively the cold dark and
baryonic matter.
The Hubble constant is $H_0=100h\;{\rm km\;s^{-1}\;Mpc^{-1}}$
with $h=0.5$, and the baryon density according to nucleosynthesis is given by
$\Omega_Bh^2=0.0125$. We will approximate the ionization history by a step
function: the universe is completely ionized before hydrogen recombination at
redshift $z_{rec}\simeq 1375$, so is after early reionization occuring at
redshift
$z_i$. Between is an universe with no optical opacity. This sudden
approximation
for the recombination and reionization processes is sufficient for the present
consideration as long as their widths in redshift are reasonably narrow. The
width will affect only calculations on small-scale anisotropies. Since we
concentrate on large-scale ($>1^o$) effect of scalar and tensor perturbations
on CMBR, we include only the dominant SW effect (Sachs \& Wolfe 1967).

To study how polarized photons propagate in the expanding universe, one need to
solve the equation of transfer for photons (Chandrasekhar 1960).  In general,
arbitrarily polarized photons are characterized by four Stokes parameters,
${\bf n}=(n_l,n_r,n_u,n_v)$, where $n=n_l+n_r$ is the distribution
function for photons with $l$ and $r$ denoting two directions
at right angle to each other.
The equation of transfer for an arbitrarily polarized photon
is governed by the collisional Boltzmann equation,
\begin{eqnarray}
\left({\partial \over \partial \eta}+{\bf e}\cdot {\partial \over \partial
\bf x} \right) {\bf n}
 &=& - {1 \over 2}{ {\partial{\bf n}} \over {\partial{\rm ln} \nu} }
              { {\partial h_{ij} }\over{\partial \eta} }e^i e^j  \nonumber\\
 &-&\sigma_T N_e a
      \left[ {\bf n} - {1 \over {4 \pi}} \int_{-1}^1 \int_0^{2\pi}
      {P(\mu,\phi,\mu^{'},\phi^{'}) {\bf n} d\mu^{'} d\phi^{'}} \right] \;,
\end{eqnarray}
\label{n}
where $\sigma_T$ is the Thomson scattering cross section, $N_e$ is
the number of free electrons per unit volume,
($\mu={\rm cos} \theta, \phi$) are the polar angles
of the propagation direction ${\bf e}$ of the photon with a comoving frequency
$\nu$,
and $P$ is the phase matrix for Thomson scattering.

The first term on the right-hand-side of equation $(\ref{n})$ is the SW effect.
For scalar modes, we have $h_{ij}=\int d{\bf k} \delta_k
e^{i{\bf k}\cdot{\bf x}}k_i k_j$,
where $\bf k$ is the wave vector and $\delta_k$ is the fluctuation amplitude
which satisfies the time evolution equation for density perturbations.
Since we are considering large-scale effect, we take
$\delta_k\propto k^2\eta^2$, which is a good approximation for long-wavelength
modes. As to tensor modes, we have $h^{\lambda}_{ij}=\int d{\bf k} h_k e^{i{\bf
k}\cdot{\bf x}}\epsilon^{\lambda}_{ij}$, where
$\epsilon^{\lambda}_{ij}$ denote the gravitational wave polarization
tensors,
$\epsilon^{+}_{ij}= \epsilon_i\epsilon_j-\epsilon^{*}_i\epsilon^{*}_j$, and
$\epsilon^{\times}_{ij}= \epsilon_i\epsilon_j^*+\epsilon^{*}_i\epsilon_j$;
where $\epsilon_i$ and $\epsilon_i^*$ are two mutually orthogonal unit vectors
perpendicular to the wave vector $\bf k$. Then, the gravitational wave
amplitude $h_k$ is governed by the equation of motion:
$\ddot h_k +2\dot a h_k/a+k^2 h_k=0$, where dot means $d/d\eta$.
The scalar and tensor power spectra are $P^{(S)}(k)\propto T(k) k^{n_s-4}$ and
$P^{(T)}(k)\propto k^{n_t-3}$ respectively, where the power indices $n_s=1$
and $n_t=0$ correspond to strict scale invariance.
For cold dark matter the scalar-mode transfer function is (Bardeen et al. 1986)
\begin{equation}
T(k)=\frac{\left[\ln(1+0.146k\eta_{eq})/(0.146k\eta_{eq})\right]^2}
{\left[1+0.242k\eta_{eq}+(k\eta_{eq})^2+(0.340k\eta_{eq})^3+(0.417k\eta_{eq})^4
\right]^{1\over 2}},
\end{equation}
where $\eta_{eq}$ is the time at which the energy density of radiation is equal
to that of matter.

The solution ${\bf n}$ for the equation of transfer assumes the form
${\bf n} ={\bf n_0} + n_0\delta {\bf n}/2$,
where ${\bf n_0}$ and $\delta {\bf n}$ are the unperturbed solution and
perturbation respectively. We expand $\delta {\bf n}=\int d{\bf k}\; {\bf n'}
e^{i{\bf k \cdot x}}$, where ${\bf n'} = \alpha {\bf a} + \beta{\bf b}$.
For the scalar-mode solution, the Stokes components $n_u$ and $n_v$ both
decouple from $n_l$ and $n_r$, and it suffices to consider only the first
two components of ${\bf n}$ with ${\bf a}=(1,1)$ and ${\bf b}=(1,-1)$.
Substituting the solution ${\bf n}$ and the Fourier expansion for $h_{ij}$ into
equation~$(\ref{n})$,
and expanding $\alpha$ and $\beta$ in terms of Legendre polynomials,
\begin{eqnarray}
\alpha(\mu) &=& \sum _{l} (2l+1) \alpha_l P_l(\mu), \nonumber \\
\beta(\mu)  &=& \sum _{l} (2l+1) \beta_l P_l(\mu),
\end{eqnarray}
we obtain two coupled differential equations for $\alpha$ and $\beta$,
\begin{eqnarray}
{\dot\alpha}+ik\mu\alpha &=& {1\over 3} T^{1\over 2}(k) k^2\eta(1+2P_2)
- q [\alpha - \alpha_0 -{1\over 2 }P_2(\alpha_2 - \beta_0 + \beta_2)],
\label{sca1} \\
{\dot\beta} + ik\mu\beta  &=& -q[\beta - {1\over 2}(1-P_2)(\beta_0
-\alpha_2  -\beta_2)],
\label{sca2}
\end{eqnarray}
where $q=\sigma_T N_e a$. These equations are equivalent to equations~(1a) and
(1b) of Bond \& Efstathiou 1984 after making the approximations that the
metric perturbations $\dot h= \dot h_{33}\propto \eta$ and the baryon velocity
$v=0$.
These approximations are valid as long as large-angular-scale calculation is
concerned.

Equations $(\ref{sca1})$ and $(\ref{sca2})$ can be casted into
a system of coupled differential equations,
\begin{eqnarray}
& &{\dot\alpha_0}= - ik\alpha_1 + {1\over 3}T^{1\over 2}(k)k^2\eta \nonumber\\
& &{\dot\alpha_1}= -{i\over 3}k(\alpha_0+2\alpha_2)-q\alpha_1 \nonumber\\
& &{\dot\alpha_2}=  -{i\over 5}k(2\alpha_1 + 3\alpha_3) +{2 \over
15}T^{1\over 2}(k)k^2\eta-{q \over 10}(9\alpha_2+\beta_0-\beta_2) \nonumber\\
& &{\dot\beta_0}= -ik\beta_1 - {q\over2}(\beta_0+\beta_2+\alpha_2) \nonumber\\
& &{\dot\beta_1}= -{i\over 3}k(\beta_0 + 2\beta_2) -q\beta_1 \nonumber\\
& &{\dot\beta_2}= -{i\over 5}k(2\beta_1 + 3\beta_3) -{q \over
10}(9\beta_2+\beta_0-\alpha_2) \nonumber\\
&&{\rm for} \;l \; \ge 3,  \nonumber \\
& &{\dot\alpha_l}= -q \alpha_l - {ik \over 2l+1 }
                           [l \alpha_{l-1} + (l+1)\alpha_{l+1}] \nonumber\\
& &{\dot\beta_l}= -q \beta_l - {ik \over 2l+1 }
                           [l \beta_{l-1} + (l+1)\beta_{l+1}] \;.
\end{eqnarray}
The coefficients $\alpha_l$ and $\beta_l$ are then solved by numerical method.

In case of tensor mode, only $n_v$ decouples, and thus we choose the basis
(Polnarev 1985): ${\bf a}= {1\over 2}(1-\mu^2){\rm cos}2\phi (1,\;1,\;0)$ and
${\bf b}= {1 \over 2} ( (1+\mu^2){\rm cos}2\phi,\;
                        -(1+\mu^2){\rm cos}2\phi,\; 4\mu {\rm sin}2\phi )$,
for the $+$ mode solution.
The $\times$ mode solution is given by the same expressions with ${\rm
cos}2\phi$ and ${\rm sin} 2\phi$ interchanged. Defining $\xi=\alpha+\beta$,
we obtain a system of coupled differential equations in a similar way,
\begin{eqnarray}
& &{\dot\xi_0}= -q\xi_0 - ik\xi_1 + {\dot h_k} \nonumber \\
& &{\dot\beta_0}= -{3\over 10}q\beta_0 - ik\beta_1
     + q ( {5 \over 7 } \beta_2 + {3 \over 35} \beta_4 -{1 \over 10} \xi_0
          +{1 \over 7} \xi_2 - {3 \over 70} \xi_4)  \nonumber \\
&&{\rm for} \;l \; \ge 1,      \nonumber \\
& &{\dot\xi_l}= -q \xi_l - {ik \over 2l+1 }
                           [l \xi_{l-1} + (l+1)\xi_{l+1}] \nonumber \\
& &{\dot\beta_l}= -q \beta_l - {ik \over 2l+1 }
                           [l \beta_{l-1} + (l+1)\beta_{l+1}] \;,
\end{eqnarray}
where the source term $\dot h_k$ is obtained by solving numerically
the equation of motion for $h_k$.

To describe the degrees of anisotropy and polarization, we
compute the power spectra for the anisotropy, $C^\alpha_l$,
and polarization, $C^\beta_l$.
To evaluate these functions, we expand the
photon fluctuation distribution function in terms of spherical harmonic
functions, i.e.,
\begin{equation}
\delta{\bf n} = \sum_{l,m} {\bf a}_{lm} Y_{lm},\quad
{\bf a}_{lm} = \int \delta{\bf n} Y^*_{lm} d\Omega.
\end{equation}
The total power spectrum is then given by
$\langle \sum_m {\bf a}_{lm}^\dagger {\bf a}_{lm}\rangle =
C^\alpha_l + C^\beta_l$, where
$\langle\rangle$ denotes the average over all observation positions in the
universe. For the scalar mode, it is straightforward to obtain (Bond \&
Efstathiou 1987)
\begin{eqnarray}
C^{\alpha(S)}_l&=&8\pi(2l+1)\int d{\bf k} P^{(S)}(k) |\alpha^{(S)}_l|^2,
\nonumber\\
C^{\beta(S)}_l&=&8\pi(2l+1)\int d{\bf k} P^{(S)}(k) |\beta^{(S)}_l|^2.
\end{eqnarray}
\label{cl}
The anisotropy and polarization power spectra for the tensor mode,
$C^{\alpha(T)}_l$ and $C^{\beta(T)}_l$ respectively, can be
deduced in a similar way (Crittenden et al. 1993; Ng \& Ng 1994).
In deriving $C^{\beta(S)}_l$ and $C^{\beta(T)}_l$, we have neglected the
rotation from the $k$-dependent basis to a fixed (laboratory) basis (see
Bond \& Efstathiou 1987 for details about the basis rotation). We have solved
our transfer equations relative to this $k$-basis and summed up all mode
contributions to the polarization power spectra in the same basis.
Rigorously, this summation would not make any sense unless one has rotated the
Stokes parameters from the $k$-basis to the fixed basis. However, for Gaussian
random perturbations, on the average, the expressions for $C^{\beta(S)}_l$ and
$C^{\beta(T)}_l$ could well represent the degree of polarization in the
laboratory (Bond \& Efstathiou 1987).
Having the power spectra, we can construct the correlation function,
\begin{equation}
C^{\alpha,\beta}(\Theta) = {1 \over {4\pi}}\sum_l C^{\alpha,\beta}_l W_l
P_l(\cos \Theta),
\end{equation}
\label{ct}
where $W_l$ is the window function for detector, and $\Theta$ is the
separation angle. In actual observations the lower end of $l$ is excluded
by limited sky coverage, whearas the high-$l$ cutoff is fixed by the finite
beam width.

\vspace{0.5 cm}
\noindent
{\centerline{\bf 3. NUMERICAL RESULTS}}
\noindent

In this section, we present the scalar and tensor contributions
to the CMBR anisotropy and polarization on large angular scales.
We separate three cases in which the anisotropy is induced by pure
scalar, pure tensor, and mixed metric perturbations respectively.
To ensure that
we capture the dominant contributions in the $k$ integration when
computing the multipoles in equation~$(\ref{cl})$,
we have investigated the $k$ dependence of certain $l$th multipole.
Roughly speaking, the dominant contribution to the $l$th
multipole comes from modes with $k\sim l$ (note that $\eta_0=1$). It is
found that for $l\leq100$,
it is sufficient to consider the contributions from modes with $k\leq 2l$,
except for the tensor-induced polarization power spectrum.
In the scalar-mode calculation, the $k$ integration must be cut off at the
point where linear perturbation theory breaks down. We thus set the
cut-off at $k_{rec}$,
where $k_{rec}$ is the wavenumber which enters the horizon during the
recombination era, since shorter-wavelength fluctuations probably have gone
nonlinear (Abbott \& Wise 1984).
We refer the interested reader to Appendix for further discussion.

Figure 1 shows the normalized CMBR anisotropy power spectra due to
scale-invariant scalar ($n_s=1$) and tensor ($n_t=0$) mode perturbations
respectively. The constant behavior of $l(l+1)C_l^{\alpha(S),(T)} / (2l+1)$
for small $l$, with a scale-invariant
spectrum in an universe with standard recombination ($z_i=0$), is evident
from the figure.  For the scalar mode, we have compared our
numerical calculation for $z_i=0$ with another alternative approach which
makes use of the SW integral formula (Sachs \& Wolfe 1967). Using this formula,
one can easily derive an analytic free-streaming solution for the anisotropy
power spectrum for $l\ge 0$ (Abbott \& Wise 1984),
\begin{eqnarray}
C^{\alpha(S)}_l&\propto& (2l+1)\int {dk\over k} {\Bigg |} \delta_l^0+
{i\over 3} k \eta_0\delta_l^1-i^l j_l[k(\eta_0-\eta_{rec})] \nonumber \\
&-& i^l k\eta_{rec} \left[\frac{l}{2l+1} j_{l-1}[k(\eta_0-\eta_{rec})]
-\frac{l+1}{2l+1} j_{l+1}[k(\eta_0-\eta_{rec})]\right] {\Bigg |}^2,
\end{eqnarray}
\label{s}
where $\eta_{rec}$ and $\eta_0$ are respectively the
recombination time and the present time. Note that we have used $\eta_{rec}$ as
the lower integration limit in the SW formula instead of the usual decoupling
time $\eta_{de}$.
Under our assumption of the ionization history, they are in fact identical.
We have plotted the power spectrum~$(\ref{s})$ represented by a long dashed
curve
in the figure. We see that, for small $l$, the result agrees very well with the
numerical result. The last term in equation~$(\ref{s})$
is essentially due to a Doppler-shift correction for the world velocity of the
source at the last scattering surface (Sachs \& Wolfe 1967). In the numerical
calculation, Thomson scatterings during the ionized stage before
photon decoupling
can effectively damp out this term. This effect is prominent mostly on
small angular scales. That is why the high-$l$ $C^{\alpha(S)}_l$'s
in the numerical result are lower than that in equation~$(\ref{s})$.
For the tensor case, similarly, we have plotted the power spectrum for $z_i=0$
caused by pure SW effect by putting the numerical solution of the evolution
equation for gravitational
waves in the SW formula (Ng \& Speliotopoulos 1994), and found good
agreement. Besides, late-time reionization ($z_i=30-90$)
reduces the CMBR temperature anisotropy on small-angular scales, and
shift the $l(l+1)C_l^{\alpha(S),(T)} / (2l+1)$ curves from scale-invariance.

Figure 2 shows the normalized CMBR anisotropy power spectra due to
tilted scalar ($n_s=0.85$) and tensor ($n_t=-0.15$) spectra respectively.
It is evident that the constancy of $l(l+1)C_l^{\alpha(S),(T)} / (2l+1)$
for small $l$ is broken, however, the relative behavior of curves with
different $z_i$ values
is similar to the scale-invariant case.  Furthermore, the normalized
magnitude of the power spectrum for large $l$ is smaller
when compared to the scale-invariant case. This can be easily explained
by referring to the
$k$ dependence of the power spectrum in equation~$(\ref{cl})$.

Figure 3 shows the polarization multipole to anisotropy multipole ratio for
scalar perturbations, $C^{\beta(S)}_l / C^{\alpha(S)}_l$, as a function of
$l$ with $n_s=1$ and $0.85$.
The ratio increases
significantly for an universe which underwent an early reionization phase.
In general, the earlier the reionization takes place, the larger is the
polarization-to-anisotropy ratio.
Note that the ratio with early reionization
has a peak around $l\sim 10-40$,
which corresponds to a few degree angular scales.  This makes the search
for large-angular-scale polarizations more interesting.  In particular,
the peak of the $z_i=90$ curve at $l \sim 20$ corresponds to
about $10\%$ polarization in the CMBR fluctuations.
Our results indicate
that the ratio is rather insensitive to the variation in $n_s$.  In fact,
there is almost no discernible difference between the $n_s=1$ and $n_s=0.85$
results. To estimate the r.m.s. polarization-to-anisotropy ratio,
$[C^\beta(0) / C^\alpha(0)]^{1/2}$, we sum $l$ from 5 up to 50 with $W_l=1$
in equation~$(\ref{ct})$, for spectra with $n_s=1, 0.85$ and different
reionization redshifts. This should correspond to typical large-scale CMBR
polarization measurements. The ratios are listed in Table~1.
In an universe with standard recombination, the polarization is much less than
$1\%$ of the anisotropy. However, the degree of polarization
with early reionization at redshift $\sim 90$ is enhanced to $6.1\%$.

Figure 4 shows the polarization multipole to anisotropy multipole ratio for
tensor perturbations, $C^{\beta(T)}_l / C^{\alpha(T)}_l$, as a function of
$l$ with $n_t=0$ and $-0.15$.
The curves are similar to the scalar case,
and the ratio is again insensitive to the variation in $n_t$.
Table~2 lists the r.m.s. large-scale polarization-to-anisotropy ratio.
The polarization with early
reionization at redshift $\sim 90$ for a scale-invariant spectrum is $9\%$ of
the anisotropy.

So far, we have considered only pure scalar- or tensor-mode
contributions to CMBR fluctuations.
In general, these fluctuations can be induced by both
scalar and tensor modes. In fact, inflation can generate both types of metric
perturbations. In inflationary models, $n_s$ is related to $n_t$ by
$n_s=n_t+1$, and there is a nearly model-independent relation between their
induced anisotropy quadrupole moment:
$C_2^{\alpha(T)}/C_2^{\alpha(S)}\simeq -7n_t$. For $n_s=0.85$ and $n_t=-0.15$,
$C_2^{\alpha(T)}\simeq C_2^{\alpha(S)}$.

In Figure 5, we have plotted the total (scalar plus tensor) polarization
multipole to anisotropy multipole ratio, $C^{\beta(S)+(T)}_l /
C^{\alpha(S)+(T)}_l$, versus $l$ for $n_s=0.85$
and $n_t=-0.15$, assuming $C_2^{\alpha(T)}=C_2^{\alpha(S)}$.  Also, in the
figure, we use short-dashed and long-dashed lines to represent the scalar
and tensor portions respectively. The tensor contribution is dominant for
almost all $l$ in the standard recombination model, however, the dominance is
taken by the scalar mode as $z_i$ increases.
Table~3 lists the r.m.s. large-scale total polarization-to-anisotropy ratio.
The polarization with early
reionization at redshift $\sim 90$ for this mixed model is $6.5\%$ of
the anisotropy.

\vspace{0.5 cm}
\noindent
{\centerline{\bf 4. COMPARISONS}}
\noindent

When comparing our result for the pure scalar mode with $z_i=0$ in Figure 3 to
Figure 7 of Bond \& Efstathiou 1987
and Figure 4 of Crittenden, Davis \& Steinhardt 1993, we find that our ratio is
about two orders of magnitude lower than theirs for
small $l$ and an order of magnitude lower for large $l$. This
may be due to the vanishing residual ionization subsequent to the hydrogen
recombination that we have assumed in our whole calculation.
In fact, we found that a residual ionization could raise the
quadrupole polarization by two orders of magnitude (Ng \& Ng 1993).
This also explains
why our r.m.s. polarization-to-anisotropy ratio ($0.06\%$, see Table~1) for the
standard
recombination model with $n_s=0.85$ pure scalar modes is much less than
Crittenden et al.'s $0.4\%$.
These discrepancies should be removed if we consider a more accurate model for
the standard thermal history of the universe. Since the degree of polarization
in the standard recombinaton model is already less than $1\%$ which is well
below the present detectable level, we will not pursue along this line and only
concentrate on the reionization model.
The $6.1\%$ result with $z_i=90$ in Table~1
is, however, consisitent with their $7.9\%$ result for the no recombination
model.

In Figure 4, the high-$l$ part of the curve with $z_i=0$ for the pure tensor
mode is
again an order of magnitude lower than the result in Figure 4 of
Crittenden, Davis \& Steinhardt 1993. But, the low-$l$ part is four
orders of magnitude lower. Once again, the residual ionization can account
for two orders of magnitude difference in the low-$l$ part. For the other two
orders of magnitude difference, we suspect that it
may be due to different definitions for the polarization power
spectrum. Unfortunately, we cannot make an explicit comparison, since their
paper did not show explicitly how to calculate the polarization spectrum.
We see from Figure 4 that the polarization to anisotropy ratio for $l=2$ is
$6.3\times 10^{-6}$.
In fact, this value is close to our earlier analytic
calculation for the tensor mode contribution to the polarization quadrupole
moment in the instantaneous recombination model with vanishing residual
ionization, which is about $9\times 10^{-6}$ (Ng \& Ng 1993).

As to the mixed-mode results, we compare our Figure 5 with Figures 1 \&
3 of Crittenden, Davis \& Steinhardt 1993. Once again, the solid curve with
$z_i=0$ in Figure 5 is well below their result probably due to the reasons that
we have mentioned above. The shape of
the solid curve with $z_i=90$ agrees quite well with their no-recombination
result.
In particular, our value for the ratio of the total polarization quadrupole
moment to the total anisotropy quadrupole moment ($\simeq 5\cdot 10^{-5}$)
agrees fairly well with their result ($\simeq 8\cdot 10^{-5}$).
Our r.m.s. total polarization-to-anisotropy ratio for the standard
recombination is $0.13\%$ (see Table~3) whereas theirs is $0.5\%$. With
reionization at redshift equal to 90, we have $6.5\%$, which is comparable to
their $7.4\%$ for the no-recombination model.
Although the general trend that the tensor contribution becomes subdominant as
$z_i$ increases is the same in both works,
the details on how the dominance taken by the scalar contribution are
different.
In our work, we observe that when $z_i=0$,
there exists a window for $l$ where the tensor mode is dominant.
As $z_i$ increases, the width of this window becomes narrow and disappears
for $z_i \geq 90$ at $l \sim 5-10$.

\vspace{0.5 cm}
\noindent
{\centerline{\bf 5. CONCLUSIONS}}

We have considered the effects of matter reionization on the
large-angular-scale
anisotropy and polarization of the cosmic microwave background radiation
(CMBR) induced by scalar and tensor metric perturbations, with scale-invariant
or tilted spectra.  The results are rather insensitive to the power indices of
fluctuation spectra.
We separate three cases in which the anisotropy is induced by pure
scalar, pure tensor, and mixed modes respectively.
It is found that the polarization is insignificant in the standard
recombination model.
But, if reionization occurs early enough, the polarization can
reach sequentially $6\%$, $9\%$, and $6.5\%$ of the anisotropy. In general, a
higher degree of polarization implies a dominant contribution from the
tensor mode or reionization at high redshift.
A $1 \%$ level sensitivity measurement could set constraint on the
reionization redshift value, as well as provide information on
whether the metric perturbation consists of a tensor component in models
with a high redshift reionization value.
For instance, if polarization is detected to be $5 \%$, then reionization
occurs at a redshift between 45 and 90, rather independent of the types
of the metric perturbation and the power spectrum index.
For the cases where $z_i \geq 80$, measurements with
$1 \%$ sensitivity can distinguish the scale-invariant tensor perturbation
from the mixed mode metric perturbation.

CMBR fluctuations have a $10\%$ level polarization on
angular scales less than $1^o$ rather model-independently. In the CDM model
with standard recombination, the predicted r.m.s. $\Delta T/T \sim 10^{-5}$ at
$1^o$. Hence, to detect small-scale polarizations would require sensitivity at
a level of $\Delta T/T \sim 10^{-6}$. In fact, this signal level can be
achieved by using new technology and instrument design (Timbie, P. T. private
communication). It is known that the universe has been reionized. Theoretical
sides also predict an early reionization in dark matter models. Our
calculations show that, if reionization did occur at redshift $\sim 90$, the
polarization to anisotropy ratio for either scalar, tensor, or mixed mode would
have a peak of height of $10\%$ around $l\sim 20$. This peak corresponds to an
angle of about $9^o$ ($\theta\sim \pi/l$). When normalized to COBE/DMR
anisotropy signals, this large-angle polarization is at a level
of $\Delta T/T \sim 10^{-6}$ and would be detectable in the near future.
Unfortunately, the early
reionization would suppress small-scale anisotropies by an order of magnitude
(Sugiyama, Silk, \& Vittorio 1993), thus making small-scale polarization
measurements elusive.
It appears that measuring CMBR polarization on both large and small angular
scales is a promising method for
determining the ionization history of the universe.

\vspace{0.5 cm}
\noindent
{\centerline{\bf APPENDIX}}

In this Appendix, we show how the $l$th multipole
depends on the range of $k$ integration.
In Figures 6 and 7, we plot for the scalar mode the spectral power spectrum
$C_l^{\alpha(S)}(k)$, where $C_l^{\alpha(S)}=\int dk\;C_l^{\alpha(S)}(k)$,
as a function of $k$ with $n_s=1$, for $z_i$ equal to 0 and 90 respectively.
In Figure 6, we also plot the dipole and quadrupole moments by using the SW
formula (equation $(\ref{s})$), and find very well agreements.
The dipole moment ($l=1)$ is
sensitive to short-wavelength fluctuations. To ensure that we do not go beyond
the linear regime, we set the cut-off $k=k_{rec}$ in our calculation.
We also investigate the $k$ dependence
of higher multipoles, and find that
the dominant contributions arise from modes with $k\leq 2l$ for the
$l$th multipole (however, it does not apply very well for small $l$).
We illustrate this dependence for $l$ equal to 2 and 20 in Figures 6 and 7.
We see that reionization does not affect much on the magnitude of the spectra
but enrich structures at their high-$k$ tails. For $l\sim 100$, the moments
will be damped out by reionization.
Similarly, in Figure 8, we plot
$C_l^{\beta(S)}(k)$ with $n_s=1$, for $l=20,40$ and $z_i=0,90$.
Obviously, reionization enhances the spectra by several orders of magnitude.
The $k\leq 2l$ rule applies very well for $l \geq 20$.  Roughly speaking, the
contributions drop by an order of magnitude when $k$ is increased to $2k$.

In Figure 9, we plot $C_l^{\alpha(T)}(k)$ for the tensor mode
as a function of $k$ with $n_t=-0.15$, for $z_i=0$ and $90$.
We notice that the $k\leq 2l$ rule applies for high multipoles.
The $l=20$ moment is damped by reionization.
In Figure 10, we plot
$C_l^{\beta(T)}(k)$ with $n_t=-0.15$, for $z_i=0$ and $90$.
Again, reionization greatly enhances the magnitude of the spectra.
For an universe with no reionization, the spectrum falls off with $k$ less
abruptly.
In this case, to capture the main contributions to the $l$th multipole, one
might need to include modes with $k\leq 10l$ in the calculation.
However, this range of $k$ is getting narrower for large $l$. At $l\sim 100$,
it is sufficient to apply the $k\leq 2l$ rule again.

\vspace{1 cm}
\noindent{\centerline{\bf ACKNOWLEDGEMENTS}}

We thank the National Center for High-Performance Computing
and the Computer Center at Academia Sinica for providing
the computer facilities, and Mr. K. Y. Hu for improving our
computer code.
This work was supported in part by the R.O.C. NSC Grant No.
NSC84-2112-M-001-024.

\noindent
\begin{table}
\caption{R.m.s. Polarization-to-Anisotropy Ratio for Scalar Mode}
\begin{tabular}{cccc}
$z_i$&$[{{C^\beta(0)}\over{C^\alpha(0)}}]^{1\over2}$
&$[{{C^\beta(0)}\over{C^\alpha(0)}}]^{1\over2}$\\
$$&$(n_s=1)$&$(n_s=0.85)$ \\ \hline
$0$&$0.0006$ &$0.00057$\\
$30$&$0.016 $  &$0.016$\\
$60$&$0.040 $  &$0.041$\\
$90$&$0.061 $  &$0.061$\\
\end{tabular}
\label{table1}
\end{table}

\begin{table}
\caption{R.m.s. Polarization-to-Anisotropy Ratio for Tensor Mode}
\begin{tabular}{cccc}
$z_i$&$[{{C^\beta(0)}\over{C^\alpha(0)}}]^{1\over2}$
&$[{{C^\beta(0)}\over{C^\alpha(0)}}]^{1\over2}$\\
$$&$(n_t=0)$&$(n_t=-0.15)$ \\ \hline
$0$&$0.0025$ &$0.0019$\\
$30$&$0.022 $  &$0.021$\\
$60$&$0.058 $  &$0.053$\\
$90$&$0.090 $  &$0.077$\\
\end{tabular}
\label{table2}
\end{table}

\begin{table}
\caption{R.m.s. Total Polarization-to-Anisotropy Ratio for Mixed
Mode} \begin{tabular}{ccc}
$z_i$&$[{{C^\beta(0)}\over{C^\alpha(0)}}]^{1\over2}$ \\
$$&$(n_s=0.85, n_t=-0.15)$ \\ \hline
$0$&$0.0013$ \\
$30$&$0.018 $ \\
$60$&$0.044 $ \\
$90$&$0.065 $ \\
\end{tabular}
\label{table3}
\end{table}

\noindent
{\bf \Large References}

\noindent
Abbott, L., \& Wise, M. 1984, Phys. Lett. 135B, 279

\noindent
Bardeen, J. M., et al. 1986, ApJ, 304, 15

\noindent
Bond, J. R., \& Efstathiou, G. 1984, ApJ, 285, L45

\noindent
Bond, J. R., \& Efstathiou, G. 1987, MNRAS, 226, 655

\noindent
Couchman, H. M. P., \& Rees, M. J. 1986, MNRAS, 221, 53

\noindent
Chandrasekhar, S. 1960, Radiative Transfer (New York: Dover)

\noindent
Crittenden, R., et al. 1993, Phys. Rev. Lett., 71, 324

\noindent
Crittenden, R., Davis, R. L., \& Steinhardt P. J. 1993, ApJ, 417, L13

\noindent
Davis, R. L., et al. 1992, Phys. Rev. Lett., 69, 1856

\noindent
Frewin, R., Polnarev, A., \& Coles, P. 1994, MNRAS, 266, L21

\noindent
Fukugita, M., \& Kawasaki, M. 1994, MNRAS, 269, 563

\noindent
Gaier, T., et al. 1992, ApJ, 398, L1

\noindent
Gunn, J. E., \& Peterson, B. A. 1965, ApJ, 142, 1633

\noindent
Harari, D., \& Zaldarriaga, M. 1993, Phys. Lett. B, 319, 96

\noindent
Kaiser, N. 1983, MNRAS, 202, 1169

\noindent
Kamionkowski, M., Spergel, D. N., \& Sugiyama, N. 1994, ApJ, 426, L57

\noindent
Krauss, L., \& White, M. 1992, Phys. Rev. Lett., 69, 869

\noindent
Liddle, A., \& Lyth, D. 1994, SUSSEX-AST 94/9-2 preprint, astro-ph/9409077

\noindent
Mather, J. C., et al. 1994, ApJ, 420, 439

\noindent
Nasel'skii, P., \& Polnarev, A. 1987, Astrofizika, 26, 543

\noindent
Ng, K. L., \& Ng, K.-W. 1993, astro-ph/9305001
to appear in Phys. Rev. D

\noindent
Ng, K. L., \& Ng, K.-W. 1994, astro-ph/9406076,
to appear in ApJ

\noindent
Ng, K.-W., \& Speliotopoulos, A. D. 1994, IP-ASTP-07-94 preprint,
astro-ph/9405043

\noindent
Polnarev, A. G. 1985, Astron. Zh., 62, 1041 [Sov. Astron., 29(6), 607]

\noindent
Rees, M. J. 1968, ApJ, 153, L1

\noindent
Sachs, R. K., \& Wolfe, A. M. 1967, ApJ, 147, 73

\noindent
Smoot, G., et al. 1992, ApJ, 396, L1

\noindent
Sugiyama, N., Silk, J., \& Vittorio, N. 1993, ApJ, 419, L1

\noindent
Tegmark, M., \& Silk, J. 1994a, CfPA-94-th-24 preprint, astro-ph/9405042

\noindent
Tegmark, M., \& Silk, J. 1994b, ApJ, 423, 529

\noindent
Tegmark, M., Silk, J., \& Blanchard, A. 1994, ApJ, 420, 484

\noindent
Vittorio, N., \& Silk, J. 1984, ApJ, 285, L39

\newpage

\noindent
{\bf Figure Captions}

Fig. 1  Normalized anisotropy power spectra.  The
solid and short-dashed curves correspond to the scale-invariant
scalar ($n_s=1$) and tensor ($n_t=0$) modes respectively, with reionization at
$z_i=0, \; 30, \;60, \; 90$. The long-dashed curve is drawn by using the SW
formula in equation~$(\ref{s})$. The long-short-dashed curve is also
drawn by using SW formula. For all curves in this figure and figures below,
$\Omega_B=0.05$ and $h=0.5$.

Fig. 2  Normalized anisotropy power spectra.  The
solid and dashed curves correspond to the tilted
scalar ($n_s=0.85$) and tensor ($n_t=-0.15$) spectra respectively, with
reionization at $z_i=0, \; 30, \;60, \; 90$.

Fig. 3  Ratio of polarization multipole to anisotropy multipole as a function
of $l$ due to scalar mode perturbations.
The solid and dashed curves correspond to the scale-invariant ($n_s=1$)
and tilted ($n_s=0.85$) cases respectively.

Fig. 4  Ratio of polarization multipole to anisotropy multipole as a function
of $l$ due to tensor mode perturbations.
The solid and dashed curves correspond to the scale-invariant ($n_t=0$)
and tilted ($n_t=-0.15$) cases respectively.

Fig. 5  Total (scalar plus tensor) polarization multipole
to anisotropy multipole ratio verus $l$, assuming
equal scalar- and tensor-induced anisotropy quadrupole moments.
The solid curves denote the total contribution. The short- and long-dashed
curves correspond to the scalar and tensor portions respectively.

Fig. 6  Spectral anisotropy power spectrum for scalar mode,
$C_l^{\alpha(S)}(k)$, as a function of $k$ with $n_s=1$ and $z_i=0$, for
$l=1,\;2,\;20$. SW denotes the results obtained by using SW formula.

Fig. 7  Spectral anisotropy power spectrum for scalar mode
with $n_s=1$ and $z_i=90$, for $l=1,\;2,\;20$.

Fig. 8  Spectral polarization power spectrum for scalar mode
$C_l^{\beta(S)}(k)$ with $n_s=1$ and $z_i=0,\;90$, for $l=20,\;40$.

Fig. 9  Spectral anisotropy power spectrum for tensor mode
with $n_t=-0.15$ and $z_i=0,\;90$, for $l=2,\;20$.

Fig. 10  Spectral polarization power spectrum for tensor mode
with $n_t=-0.15$ and $z_i=0,\;90$, for $l=2,\;20$.

\end{document}